*Title:*

# Persistent and anti-persistent pattern in stride-to-stride variability of treadmill walking: influence of rhythmic auditory cueing

*Authors:*

Philippe Terrier[1,2,§], Olivier Dériaz[1,2]

*Author contributions:*

P.T. designed and performed the experiment, analyzed the data and wrote the article. O.D. supervised the study, gave conceptual advice, and edited the manuscript.

*Affiliation:*

1 IRR, Institut de Recherche en Réadaptation, Sion, Switzerland

2 Clinique romande de réadaptation SuvaCare, Sion, Switzerland

*Running head:*

Effects of auditory cueing on gait dynamics

*Contact information:*

[§] Corresponding author:

Dr Philippe Terrier, PhD

Recherche médicale

Av. Gd-Champsec 90

1951 Sion

SWITZERLAND

Tél. +41(0)27 603 20 70

E-mail : philippe.terrier@crr-suva.ch

**Abstract**

It has been observed that long time series of Stride Time (ST), Stride Length (SL) and Stride Speed (SS=SL/ST) exhibited statistical persistence (long-range auto-correlation) in overground walking. Rhythmic auditory cueing induced anti-persistent (or anti-correlated) pattern in ST series, while SL and SS remained persistent. On the other hand, it has been shown that SS became anti-persistent in treadmill walking, while ST and SL remained persistent. The aim of this study was to analyze the effect of the combination of treadmill walking (imposed speed) and auditory cueing (imposed cadence) on gait dynamics. Twenty middle-aged subjects performed 6 x 5min walking trials at various imposed speeds on an instrumented treadmill. Freely-chosen walking cadences were measured during the first three trials, and then imposed accordingly in the last three trials by using a metronome. Detrended fluctuation analysis (DFA) was performed on the times series of ST, SL, and SS. Treadmill induced anti-persistent dynamics in the time series of SS, but preserved the persistence of ST and SL. On the contrary, all the three parameters were anti-persistent under dual-constraints condition. Anti-persistent dynamics may be related to a tighter control: deviations are followed by a rapid over-correction, what produces oscillations around target values. Under single constraint condition, while SS is tightly regulated in order to follow the treadmill speed, redundancy between ST and SL would likely allow persistent pattern to occur. Conversely, under dual constraint conditions, the absence of redundancy among SL, ST and SS would explain the generalized anti-persistent pattern.

Highlights

- We studied statistical persistence of Stride Time (ST), Stride Length (SL) and Stride Speed (SS)
- When walking on a treadmill, ST and SL are persistent, while SS is anti-persistent
- When rhythmic auditory cueing (metronome) is added, SL, ST and SS are anti-persistent
- A model based on the redundancy theory in the movement control is proposed
- The goal function constraining both ST and SS lead to a concomitant anti-persistence of SL, ST and SS

1. **Introduction**

Human walking associates the goal of forward progression with the need for a safe motion. A continuous balance control is required to prevent falling. In addition, an efficient postural control keeps the body upright and a fine motor control ensure a safe foot clearance and a soft heel contact (Winter, Patla, & Frank, 1990). These tasks result from complex dynamic sensorimotor interactions. At spinal level, neural circuitry (Central Pattern Generator) generates the basic locomotor pattern, under the control of various descending pathways (Rossignol, Dubuc, & Gossard, 2006). Visual, auditory and vestibular sensory inputs, as well as afferences from muscles and skin, constitute inputs for feedback and feedforward mechanisms that constantly adapt locomotion to the environment. This multi-level neural control system produces a highly consistent walking pattern. In steady conditions, kinetics, kinematics and muscular activity appear to remain relatively constant from one stride to the next (Hausdorff, 2007). However, small residual stride-to-stride fluctuations occur as the result of internal, neuromuscular noise and small adaptations to the changing environment (such as, for instance, irregularities of the walking surface).

Because an healthy gait mainly consists in maintaining steady speed, Stride Length (SL), Stride Time (ST) and the resulting Stride Speed (SS=SL/ST) can be viewed as "final output" of the multi-dimensional neuromuscular control system, which integrates the result of the different sensory-motor processes. Thus, the analysis of SL, ST and SS could theoretically provide insight into the neurophysiological organization of the motor control and into the regulation of the entire locomotor system (Hausdorff, 2007). For instance, it has been observed that the ratio between step length (SL) and step frequency (SF), or Walk Ratio, is constant over a large range of walking speed (Terrier & Schutz, 2003). In other words, in a physiological range of speeds around preferred walking speed, a linear relationship exists between SL and SF. It seems that this combination of ST and SL is related to energy expenditure optimization (Terrier & Schutz, 2003; Zarrugh, Todd, & Ralston, 1974). This automated / involuntary adaptation could result from the interaction from lower-level pattern generating mechanism with the dynamical movement context (Zijlstra, Rutgers, Hof, & Van Weerden, 1995); furthermore, it as been suggested that interactions between basal ganglia and supplementary motor area could provide the correct stride length and cadence for optimal efficiency (Egerton, Danoudis, Huxham, & Iansek, 2011). Conversely, when a step-by-step control is needed, either in term of ST modulations (i.e. controlling cadence) or SL modulations (i.e. controlling step length), specific supraspinal mechanisms have been suspected, (Zijlstra et al., 1995), possibly implying premotor cortex and supplementary motor area (Halsband, Ito, Tanji, & Freund, 1993).

The synchronization of body movements to external rhythm (auditory-motor coordination) is a remarkable ability of the human brain (Repp, 2005; Zatorre, Chen, & Penhune, 2007). Step time modulations driven by rhythmic auditory cueing have been studied in the context of different clinical disorders, such as head injuries, Parkinson's disease or stroke (Lim et al., 2005). It can induce a substantial beneficial effect on gait performance (Lim et al., 2005; Nieuwboer et al., 2007). Rhythmic auditory stimuli may be efficient because they would stimulate intact auditory-motor system. For instance, in Parkinson's disease, it is thought that this strategy enables the use of a functional auditory-motor circuit instead of relying on impaired motor control implying affected basal ganglia.

Understanding stride-to-stride control requires quantifying not only average magnitudes of variations across strides (for instance Standard Deviation, SD), but also the specific temporal sequencing of those variations. In the mid 90's, it has been suggested that successive stride durations (ie. time series of ST) during walking presented a typical structure over time, characterized by the presence of statistical persistence (Delignieres & Torre, 2009; Hausdorff, Peng, Ladin, Wei, & Goldberger, 1995; Hausdorff et al., 1996). In addition to ST, Terrier et al. described persistent patterns in SL and SS during long duration overground walking (Terrier, Turner, & Schutz, 2005). "Persistence" means that deviations are statistically more likely to be followed by subsequent deviations in the same direction (i.e. persist across subsequent data points). Conversely, "Anti-persistence" means that deviations in one direction are statistically more likely to be followed by subsequent deviations in the opposite direction.

As in human locomotion, it has been observed in many other physiological time series that the current value possesses the memory of preceding values. This phenomenon –referred alternatively to as long-range correlations, long-term memory, long-range dependence, fractal process or 1/f noise– has been identified in a number of systems and situations, such as heartbeat time series or in the time intervals produced in finger tapping (Diniz et al., 2011). An unified theory explaining this widespread characteristic in physiological processes is still to be built (Diniz et al., 2011). One explanation could be that fractal processes are natural outcome of complex self-organized systems, emerging from cooperation between their components acting at different space or time scales. Furthermore, it has been proposed that this fractal-like structure was related to the high adaptability of healthy individuals (Goldberger et al., 2002). Conversely, breakdown of the fractal-like structure (i.e. the emergence of more uncorrelated time series) has been associated with various diseases (Goldberger et al., 2002; Hausdorff, Cudkowicz, Firtion, Wei, &

Goldberger, 1998; Hausdorff et al., 1997; Peng, Havlin et al., 1995). It was claimed that the loss of statistical persistence would be related to decreased adaptability of neural structures and looser cortical control.

Concerning human gait, methodological issues have questioned the existence of true long-range correlations in ST time series (Maraun, Rust, & Timmer, 2004). On the other hand, a study confirmed the presence of long-range correlations in gait signals by using an alternative statistical approach (Delignieres & Torre, 2009). Moreover, it has been suggested that the purported long-range correlations may not require complex central nervous system control mechanisms, but might be due instead to the inherent biomechanical structure of the system (Gates, Su, & Dingwell, 2007). The paradigm of statistical persistence as an index of healthy system –and its corollary associating uncorrelated pattern with diseases— has also been challenged: indeed, when healthy individuals walked following the cadence of a metronome, ST exhibited either uncorrelated (Hausdorff et al., 1996) or anti-persistent pattern (Delignieres & Torre, 2009; Terrier et al., 2005). It has also been suggested that cautious gait could influence persistent pattern (Herman, Giladi, Gurevich, & Hausdorff, 2005). Furthermore, by analyzing the persistence of ST, SL and SS in treadmill walking, a recent study (Dingwell & Cusumano; Dingwell, John, & Cusumano) found not only that SL and ST exhibited strong statistical persistence, but also that SS exhibited anti-persistent dynamics. The authors suggested that statistical anti-persistence emerges from stride-to-stride fluctuations because of increased central control. Deviations in SS are followed by rapid corrections in order to follow the treadmill speed. However, a slight but repeated over-correction lead to an anti-persistent dynamics.

In view of the results in the literature, an evident research question arises: what happens when treadmill and metronome are used simultaneously? The objective of the present study was therefore to analyze the interactions between two constraints: treadmill, which imposes a constant speed, and rhythmic auditory cueing, which imposes a constant cadence. We assessed, in healthy individuals, statistical persistence (or anti-persistence) in times series of SL, ST and SS by using Detrended Fluctuation Analysis (DFA) at preferred speed as well as at low and high speed, with and without rhythmic auditory cueing. We tested the hypothesis, that only gait variables directly relevant to achieving the task goal (speed and cadence constraints) would exhibit statistical anti-persistence.

2. Methods

*2.1. Participants.*

Twenty healthy subjects (10 females, 10 males) participated in the study. Based on preliminary interviews, we ensured that they did not exhibit any orthopedic problems or other health issues that could influence their gait. They were recruited in Sion area (South-west Switzerland). The participants' characteristics were (mean (SD): age 36yr (11), body mass 71kg (15), and height 171cm (9). The average Body Mass Index (BMI) was 24 (4), what is close to the average BMI of the Swiss population (24.9, (Volken, Schaffert, & Ruesch, 2011)). The study was approved by the local ethics committee (Commission d'éthique du Valais).

*2.2. Experimental procedure*

Their Preferred Walking Speed (PWS) on the treadmill was determined as follows: a) the treadmill speed was progressively increased by 0.2km/h from low speed (2km/h) until comfortable pace; b) the procedure is then repeated by progressively decreasing speed from high speed (5.5km/h) to comfortable pace. PWS was defined as the average of both values. The protocol design consisted in measuring three different speeds with and without auditory cueing (metronome walking). In more details, the speeds imposed to the subjects were: PWS, 0.7 x PWS (low speed) and 1.3 x PWS (high speed). The sequence of the trials with three different speeds was randomly designed. Each walking trials lasted 5m30: 30sec of habituation to the speed, and 5min of measurement. After a 30min resting period, the trials with "metronome" condition were performed at the same speeds as the first 3 trials (random sequence between speeds). The imposed cadences were the preferred cadences, which were measured during the first session: as a result, the subjects had no difficulty to follow the imposed rhythm. The subjects were told to "synchronize their heel strikes with the "tack" of the electronic metronome".

*2.3. Instrument*

The instrumented treadmill was a FDM-TDL (Scheinworks/Zebris, Schein, Germany). The size of the walking surface was 150cm x 50cm. Two security rails run along the first front quarter of the treadmill. The sensor surface contained 7168 (128x56) pressure sensors on a 108.4 x 47.4 cm grid (1.4 sensors per cm$^2$). Pressure data were sampled at 100Hz. We obtained a "movie" of the feet pressure on the treadmill belt. The raw data consisted for each trial in 30'000 frames of 56x128 points: they were exported for subsequent analysis with Matlab (Mathworks, USA).

*2.4. Time series*

The time series of ST, SL and SS were obtained by specifically detecting the heel strikes in the frames, and hence calculating time and distance between consecutive points. More precisely, heel strike was detected when a frame with one foot print (end of swing phase) is followed by a frame with two foot prints (beginning of stance phase). The position and the time of each heel strike allowed us to obtain the length and duration of each step. We computed ST, SL and SS (=SL/ST) for both right and left sides. The final results (i.e. CVs and scaling exponents, see below) were computed separately for each side and then averaged. The variability in the time series was assessed by computing Coefficient of Variation (CV), i.e. the Standard Deviation (SD) of ST, SL and SS divided by the mean. In order to highlight how step length and step frequency (inverse of step time) varied together with speed, we computed the Walk Ratio (WR). WR is the ratio between step length and step rate (cadence). It is thought that this ratio is constant over a large range of walking speeds (Terrier & Schutz, 2003).

*2.5. Detrended Fluctuation Analysis*

DFA is based on a classical root-mean square analysis of a random walk, but is specifically designed to be less likely affected by non-stationarities. More details of the methodology can be found in the literature (Dingwell et al., 2010; Hausdorff et al., 1995; Peng, Buldyrev et al., 1995; Terrier & Deriaz, 2011; Terrier et al., 2005). The detection of fractal properties by using this method has been questioned. It seems that short-range dependence in time series can lead to erroneous interpretation (false positive) (Maraun et al., 2004; Wagenmakers, Farrell, & Ratcliff, 2004). In order to not over-interpret DFA results, we will use throughout the present article the notion of persistence rather than long-range correlations. Indeed, statistical persistence characterizes both short-range and long-range autocorralated process.

The integrated time series of length N is divided into boxes of equal length, n. In each box of length n, a least squares line is fit to the data (representing the trend in that box). The y coordinate of the straight line segments is denoted by $y_n(k)$. Next, the integrated time series, $y(k)$, was detrended, by subtracting the local trend, $y_n(k)$, in each box. The root-mean-square fluctuation of this integrated and detrended time series is calculated by

$$F(n) = \sqrt{\frac{1}{N} \sum_{k=1}^{N} [y(k) - y_n(k)]^2} \qquad (1)$$

This computation is repeated over increasing box sizes (from 5 to 120) to assess the relationship between F(n), the average fluctuation, and the box size, n. We preliminary tested different ranges of box sizes (5 to N/4, 5 to N/2, 5 to 120) with no significant change on the final results. The fluctuations can be characterized by the scaling exponent α, which is the slope of the line relating log F(n) to log(n) (F(n) ~ $n^{α}$). The signal exhibit a statistical persistent pattern when α lies between 0.5 and 1, and an anti-persistent one when α<0.5. The limit of 0.5 is the evidence for an uncorrelated white noise.

*2.6. Statistics*

The range of the data was depicted by using notched box-plots (fig. 1-3). The boxes have lines at the lower quartile, median, and upper quartile values. The whiskers are lines extending from each end of the boxes to show the extent of the rest of the data, in the range of 1.5 times the upper an lower quartiles. Outliers are data with values beyond this limit. The notches represent an estimate of the uncertainty about the medians for box-to-box comparison. Boxes whose notches do not overlap indicate that the medians differ at the 5% significance level. The size of the effect induced by rhythmic auditory cueing was estimated by computing the Hedge's g, which is a modified version of the Cohen's d for inferential measure (Nakagawa & Cuthill, 2007). Confidence intervals (CI) were ±1.96 times the asymptotic estimates of the standard error (SE) of g. The arbitrary limit of 0.8 was uses to delineate large effect sizes, as defined by Cohen.

*2.7. Surrogate data analysis.*

The principle of surrogate data analysis is to create –by shuffling– numerous alternative time series from the original one. It has been developed as an approach for evaluating the statistical significance of evidence for nonlinearity in time series (Theiler, Eubank, Longtin, Galdrikian, & Farmer, 1992). Surrogate data testing attempts to find the "least interesting" explanation that cannot ruled out based on the data (Schreiber & Schmitz, 2000). In the context of DFA, this method is routinely used to exclude that the analyzed time series is the result of a random process with no correlation between successive points: the comparison of different shuffled versions of the times series with the original one helps to determine how well random processes might account for the experimental data. In the present study, we followed the method of Dingwell et al. (Dingwell & Cusumano, 2010; Dingwell et al., 2010) in order to present comparable results. Please refer to these articles to obtain more details. Four types of surrogate time series were analyzed; for each type and each original time series 20 surrogate series were generated:

1) *Randomly and independently shuffled surrogates.* For each trial, original ST and SL time series were shuffled in a random order. All effects of temporal order and any dependence between ST and SL were lost.

2) *Paired randomly shuffled surrogates.* For each trial, surrogates were generated simultaneously by randomly shuffling both SL and ST in exactly the same way. As a result, a potential cross-correlation between ST and SL was preserved, although temporal order was lost.

3) *Independently phase-randomized surrogates.* These surrogates required to shuffle the phase of the original time series in the frequency domain, independently for ST and SL. Consequently, temporal dependence across consecutive strides was preserved, but potential cross-correlations between ST and SL were lost.

4) *Paired phase-randomized surrogates.* These surrogates required to shuffle the phase of the original time series in the frequency domain, the same way for both ST and SL. This method preserved auto-correlation inside-, and cross-correlation between-, time series.

Each surrogate set was computed by ensuring that the maximal net cumulative distance virtually walked did not exceed the treadmill belt limits. Therefore, these "virtual" surrogates mimicked original, "human", time series. Surrogate SS time series was computed from surrogates of ST and SL (SS=SL/ST). DFA was applied for each surrogate, and then the resulting scaling exponent was averaged (N=20 surrogates per trial). Standard deviation (N=20 surrogates) was also assessed. In order to evaluate the extent of the surrogates across each trial (N=20) and among subjects (N=20), we computed average SD.

3. Results

The figure 1 shows the range of speeds chosen by the participants for each condition and the corresponding Walk Ratio (i.e. the ratio between step length and cadence). WR is constant across speeds, with a slight tendency of higher WR at low speed condition. Figure 2 shows the extent of the data for stride-to-stride variability, expressed as Coefficient of Variation (CV, i.e. Standard Deviation / mean * 100). Both conditions are presented: treadmill only (top) and treadmill combined with auditory cueing (metronome, middle). The lower panels show the Effect Size (ES) of rhythmic auditory cueing. Auditory cueing strongly reduced the variability of ST at low speed (ES=-1.1). The effect on ST is more attenuated at preferred walking speed (PWS, ES=-0.6) and not significant at high speed (ES=-0.18). On the other hand, auditory cueing seemed to have lower impact on SL (ES=-0.81, -0.46, -0.23 for low, PWS, and high speed, respectively), and only a small effect on SS (ES=-0.45, -0.33, -0.26 for low, PWS, and high speed, respectively).

The figure 3 presents the results of the Detrended Fluctuation Analysis (DFA). When the subjects walked on the treadmill without auditory cueing, both ST and SL exhibited statistical persistence. On the contrary, SS was strongly anti-persistent. When the rhythmic auditory cueing was added, all the three parameters were anti-persistent. The ES confirms that the synchronization with the metronome had a huge effect (ES larger than 2) on both ST and SL, but a non-significant effect on SS.

As expected, by eliminating all effects of temporal order, randomly shuffled surrogates did not replicate the original data (fig 4.): ST and SL, as well the SS combination (SS=SL/ST), yielded uncorrelated time series ($\alpha$=0.5) for both conditions (treadmill only and treadmill+metronome). The phase-randomized surrogates (fig. 5) preserved auto-correlations, but not the same temporal order as the original series. As a result, the original statistical persistence is preserved in the surrogate series of ST and SL. Similarly, statistical anti-persistence of ST and SL was also preserved in the metronome condition. Conversely, when combined (SS=SL/ST), the independently shuffled surrogates did not replicate the experimental results: a persistent dynamics was observed (fig 5, triangles). On the other hand, the paired phase-randomized surrogates, which randomized ST and SL in the same manner, yielded strong anti-persistent SS, replicating the original data (fig 5, squares). This suggests that an underlying cross-correlation existed in the original SL and ST series. In the metronome condition, the independently randomized surrogates of SS were slightly more uncorrelated (i.e. $\alpha$ closer to 0.5) than their paired randomized counterpart, especially at PWS and high speed, what corroborates the underlying cross-correlated pattern.

4. Discussion

By using an instrumented treadmill, we measured the length and duration of gait cycles at 3 walking speeds (slow, preferred, fast) and under two conditions (without and with rhythmic auditory cueing). We observed that auditory cueing induced a decrease in the variability of ST and SL (lower CV), especially at low speed. Treadmill induced anti-persistent dynamics in the time series of SS, but preserved the persistence of ST and SL. On the contrary, all the three parameters were anti-persistent under dual-constraints condition.

Several methodologies have been used to measure long-term time series of ST, such as pressure-sensitive foot-switches (Hausdorff et al., 1998; Hausdorff et al., 1997; Hausdorff et al., 1995), or video analysis (Jordan, Challis, Cusumano, & Newell, 2009). Few studies attempted to simultaneously measure time series of SL and ST by using instrumented treadmill (embedded force plates) (Jordan, Challis, & Newell, 2007), video analysis (Dingwell & Cusumano; Dingwell et al.) or high accuracy GPS (Terrier et al., 2005). As far as we are aware, the present study was the first attempt to collect long time series of SL, ST and SS by using a treadmill, instrumented with foot-pressure sensors aimed at dynamic plantar pressure assessment. The main advantage of this technique is the easiness of the measurement procedure (no marker, no calibration). Because we obtained comparable results as others (see below), temporal (0.01s) and spatial (0.85cm) resolution seemed sufficient to report stride-to-stride variability and fluctuation pattern.

The objectives of the present study did not include the thorough analysis of speed effects. The design was build in order to ensure that the potential auditory cueing effect was not limited to PWS, but was also acting at slow and high speeds: as presented in figure 3, this was the case. It has been observed that scaling exponents exhibited quadratic dependence (U-shaped form) as a function of speed (Jordan et al., 2007). Such dependence cannot be satisfactorily analyzed with only 3 different speeds. Globally, our results confirm that low speed induces higher stride-to-stride variability (Jordan et al., 2007; Kang & Dingwell, 2008; Terrier & Schutz, 2003). In addition, increasing speed seemed to produce slightly less correlated pattern ($\alpha$ closer to 0.5). Further studies are needed to better address the speed issues.

Rhythmic auditory cueing reduced stride to stride variability of ST, especially at low speed. The effect is less strong for SL and low for SS. In contrast, in a previous "overground walking" study (Terrier et al., 2005), at PWS, no significant differences were observed between metronome and free-walking condition concerning stride-to-stride variability of SL, ST and SS. It has been reported that low speed could induce higher Walk Ratio (WR) in overground walking for a majority of individuals (Terrier & Schutz, 2003). The results (fig. 1, bottom) show the same trend. It is likely that slow walking (70% PWS) induced a specific spatial and temporal adaptation of the gait (higher WR) that was "un-natural" for many subjects, what induced a larger variability. By giving a landmark, the metronome stabilized the stride-to-stride variations (-33% CV reduction, cf. fig. 2). This "landmark" hypothesis could also be applied to SS: this parameter exhibit a lower CV, with only a slight reduction with auditory cueing, because motor control used the treadmill speed as reference.

We confirm that treadmill specifically induces an anti-persistent pattern in SS stride-to-stride fluctuations (fig. 3), with ST and SL remaining persistent. Several studies have been conducted to explain statistical persistence and long range correlations (scaling exponent $\alpha \cong 0.8$) in stride-to-stride fluctuations of healthy gait (Dingwell et al., 2010; Diniz et al., 2011; Goldberger et al., 2002; Hausdorff, 2007). Simple explicative stochastic models have been proposed (Ashkenazy, Hausdorff, Ivanov, & Stanley, 2002; West & Scafetta, 2003). In contrast, few studies described low scaling exponent and anti-persistent fluctuations as normal output of the motor control while walking (Delignieres & Torre, 2009; Dingwell & Cusumano, 2010; Terrier et al., 2005). The recent results of Dingwell et al. (Dingwell & Cusumano, 2010; Dingwell et al., 2010) shed a new light upon the interpretation of anti-persistent dynamics in gait parameters. According to these authors, a greater stride-to-stride control ("cautious gait") could lead to a more uncorrelated pattern (lower scaling exponent) in gait parameters (Dingwell & Cusumano, 2010). By analyzing the anti-persistent pattern of SS during treadmill walking, they showed that a simple stochastic model, which sub-optimally over-corrected the stride-to-stride deviations, could reproduce the human results. In other words, it seems that when a tight control of a gait variable is needed (i.e. more voluntary/conscious motor control), a deviation in this variable is followed by a rapid correction. However, the correction is not optimal, but slightly over-estimated by the motor control. As a result, the controlled gait variable tends to "oscillate" around the target value, what produces the observed anti-persistent pattern. The exact neural mechanisms behind such a sub-optimal control remain to be further investigated. The phenomenon could be explained by corrective process, with respect to an internal model (Repp, 2005; Torre & Delignieres, 2008). It could be hypothesized that a conflict arises between internal models that feed the motor output forward and the sensory inputs that feed the motor control back. Such discrepancies between central representation and peripheral actuation have been described in finger tapping (asynchrony issue (Aschersleben, 2002)). However, it is worth noting that other authors have suggested that purely non-cognitive processes, implying coupling between oscillators, with no internal model or correction, could generate anti-persistent pattern (Delignieres & Torre, 2009; Torre, Balasubramaniam, & Delignieres, 2010).

The main new finding of the present study is the generalized anti-persistent pattern (affecting ST, SL and SS) under dual constraints condition (treadmill + metronome). This contradicts our working hypothesis, which was that only the controlled parameters (i.e ST and SS) should become anti-persistent. On the contrary, we observed that SL also exhibited strong anti-persistent dynamics, although no direct constraint was imposed on this gait parameter. An explanation could be based on the recent theories of motor coordination, such as the uncontrolled manifold approach (Latash, Scholz, & Schoner, 2002) or the minimal intervention principle (Todorov, 2004). These authors make the link between motor variability and motor redundancy. Motor redundancy refers to the fact that the motor control has numerous alternatives to perform a given task. It is thought that motor control allows high variability (more freedom) to parameters, which do no affect the desired value of the variable. On the contrary, it restricts the variability of parameters that are essentials for achieving the task. In the context of redundancy in gait control, Dingwell et al. (Dingwell et al., 2010) developed a new theory explaining persistent and anti-persistent pattern in treadmill walking. They proposed that a "Goal Equivalent Manifold" (GEM) exists between SL and ST. In order to maintain a constant SS, a walking individual could choose an infinite combination of ST and SL (SS=SL/ST): in other words, in a diagram of ST vs. SL, the goal function (GEM, maintaining constant SS) is a diagonal line, along which motor control must keep, stride after stride, the [ST SL] pairs. While SS is tightly regulated in order to follow the treadmill belt, deviations in both ST and SL could be allowed to persist with no consequences: deviations in either variable can be "cancelled out" by concomitant changes in the other (Dingwell & Cusumano, 2010). In other words, the degrees of freedom are sufficient to allow statistical persistence to occur, because there is a redundancy between SL and ST in achieving a given SS. It is worth noting that treadmill by itself significantly lower ST scaling exponent as compared to overground walking (Terrier & Deriaz, 2011), what could be a sign of the "hidden" cross-regulation of ST and SL producing anti-persistent SS. In overground walking, it has been shown that auditory cueing specifically affects ST persistence, with no effect on SL and SS (Terrier et al., 2005). In this case, the GEM for walking at constant ST would be a vertical line in the [ST SL] diagram, and hence there is a redundancy between SL and SS in achieving the imposed ST. When treadmill and auditory cueing are combined, an extra degree of freedom is lost. Motor control must achieve two GEM simultaneously: the diagonal line in the [ST SL] diagram (imposed by the constraint on SS) and the vertical line (imposed by the constraint on ST). Thus, there is no longer any remaining redundancy between SL, ST, and SS. To achieve both goals simultaneously, subjects must correct deviations in all directions away from the single goal point in the [ST SL] diagram, including deviations in unconstrained SL. This hypothesis of simultaneous [ST SL] control is supported by the results of the phase-randomized surrogate testing, which show that the cross-correlated model fits better to the humans' results than the uncorrelated one (fig.5).

In summary, the results of the present study, taken together with the results of previous studies (Dingwell & Cusumano, 2010; Dingwell et al., 2010; Terrier et al., 2005; Zijlstra et al., 1995), suggest that two modalities of gait control co-exist: 1) a more automated/unconscious mode, which produces persistent, fractal-like pattern across numerous successive strides, what is likely related to the redundancy among the gait parameters to achieve steady gate; 2) a more voluntary/conscious mode relying upon fast over-correction of deviations in the controlled variable, which produces anti-persistent pattern among successive strides. In addition, the combination of two constraints (time and speed) induces anti-persistence in the three gait variables (ST, SL and SS), what is likely the result of the cross-regulation of SL and ST and the absence of redundancy among the gait parameters to achieve the dual goal function.

**Acknowledgments**

The authors thank M. Philippe Kaesermann for the loan of the instrumented treadmill, and the anonymous reviewers for their constructive remarks. The study was supported by the Swiss accident insurance company SUVA, which is an independent, non-profit company under public law. The IRR (Institute for Research in Rehabilitation) is supported by the State of Valais and the City of Sion.

FIGURE CAPTIONS

Fig. 1.

**Treadmill speed and Walk Ratio.** *Twenty healthy subjects walked 3x5min. on a motorized treadmill. Selected speeds were Preferred Walking Speed (PWS), 0.7 times PWS and 1.3 PWS. Upper panel shows the speeds performed by the subjects for the three different levels. Lower panel show the Walk Ratio, i.e. the ratio between step length (in meter) and cadence (step per second). Extent of the data (N=20) is presented with notched boxplots. Printed values are mean(SD).*

Fig. 2.

**Stride-to-stride variability.** *Twenty healthy subjects walked 3x5min. on a motorized treadmill without (top) and with (middle) rhythmic auditory cueing (metronome) at their preferred cadence for the given speed. Selected speeds were Preferred Walking Speed (PWS), 0.7 times PWS and 1.3 PWS. Stride-to-stride variability of Stride Time (ST), Stride Length (SL) and Stride Speed (SS) is expressed as CV (Coefficient of Variation, SD / mean\*100). Extent of the data (N=20) is presented with notched boxplots. Printed values are mean(SD). Bottom panels show the effect size (Hedge's g, variant of Cohen's d) of the auditory cueing, i.e. the mean difference normalized by SD. Vertical lines are the 95% confidence intervals for the effect size estimations.*

Fig. 3.

**Statistical persistence.** *Twenty healthy subjects walked 3x5min. on a motorized treadmill without (top) and with (middle) rhythmic auditory cueing (metronome) at their preferred cadence for the given speed. Selected speeds were Preferred Walking Speed (PWS), 0.7 times PWS and 1.3 PWS. Detrended Fluctuation Analysis (DFA) was performed on the time series of Stride Time (ST), Stride Length (SL) and Stride Speed (SS). The scaling exponent α is an estimator of the statistical persistence present in the time series: persistence => α>0.5 ; anti-persistence => α<0.5. Extent of the data (N=20) is presented with notched boxplots. Bottom panels show the effect size (Hedge's g, variant of Cohen's d) of the auditory cueing, i.e. the mean difference normalized by SD. Vertical lines are the 95% confidence intervals for the effect size estimations.*

Fig. 4.

***Surrogate data tests: random shuffling.*** *Circles show the same data as in fig 3 (mean scaling exponents). The range of the inter-individual variability is represented by Standard Deviation ( ±SD, vertical lines, N=20 subjects)). Triangles are the mean results for randomly and independently shuffled surrogates (N=20 surrogates / 20 subjects). Squares are the mean results for the paired randomly shuffled surrogates (N=20 surrogates / 20 subjects). Vertical lines are the average range of variability among surrogates and subjects, i.e. mean SD (N=20 surrogates, averaged across 20 subjects).*

Fig. 5.

***Surrogate data tests: phase-randomized.*** *Circles show the same data as in fig 3 (mean scaling exponents). The range of the inter-individual variability is represented by Standard Deviation (± SD, vertical lines, N=20 subjects)). Triangles are the mean results for independently phase-randomized surrogates (N=20 surrogates / 20 subjects). Squares are the mean results for the paired phase-randomized surrogates (N=20 surrogates / 20 subjects). Vertical lines are the average range of variability among surrogates and subjects, i.e. mean SD (N=20 surrogates, averaged across 20 subjects).*

*Figure 1*

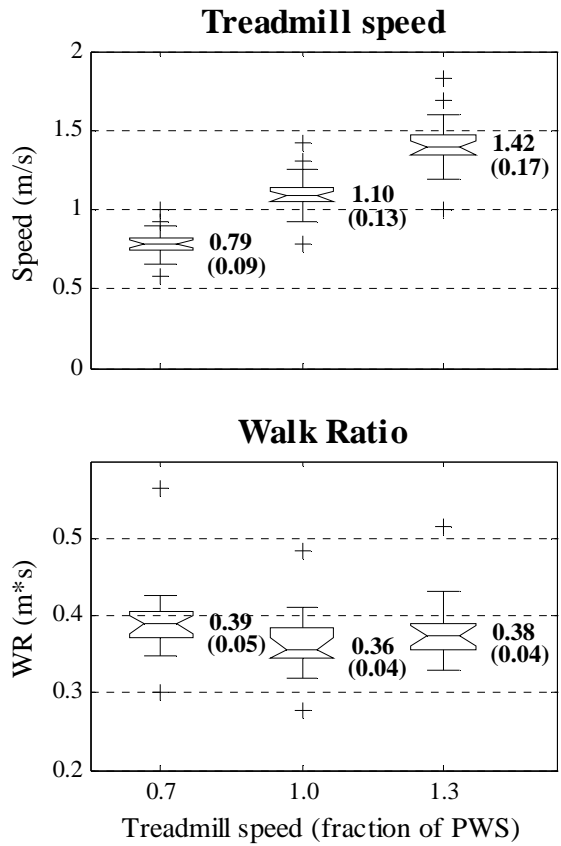

*Figure 2*

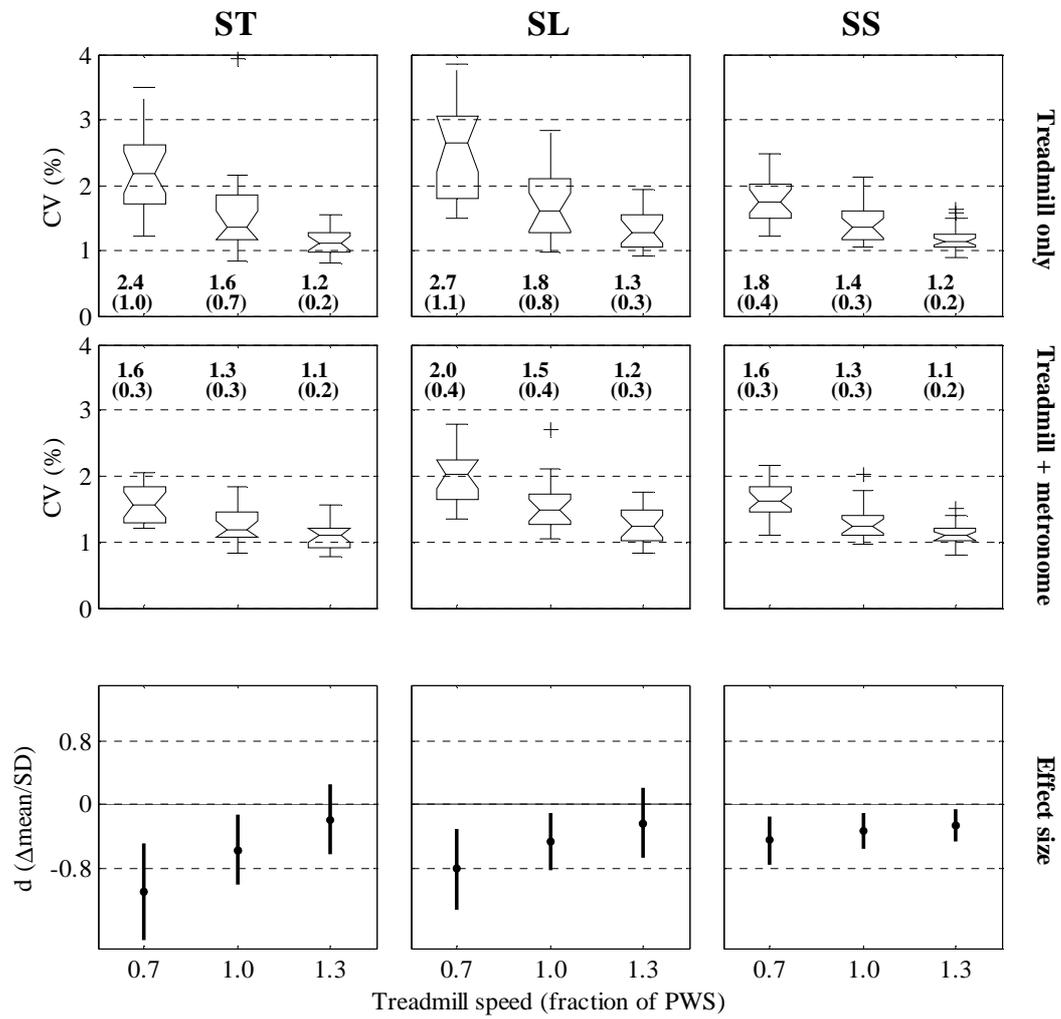

*Figure 3*

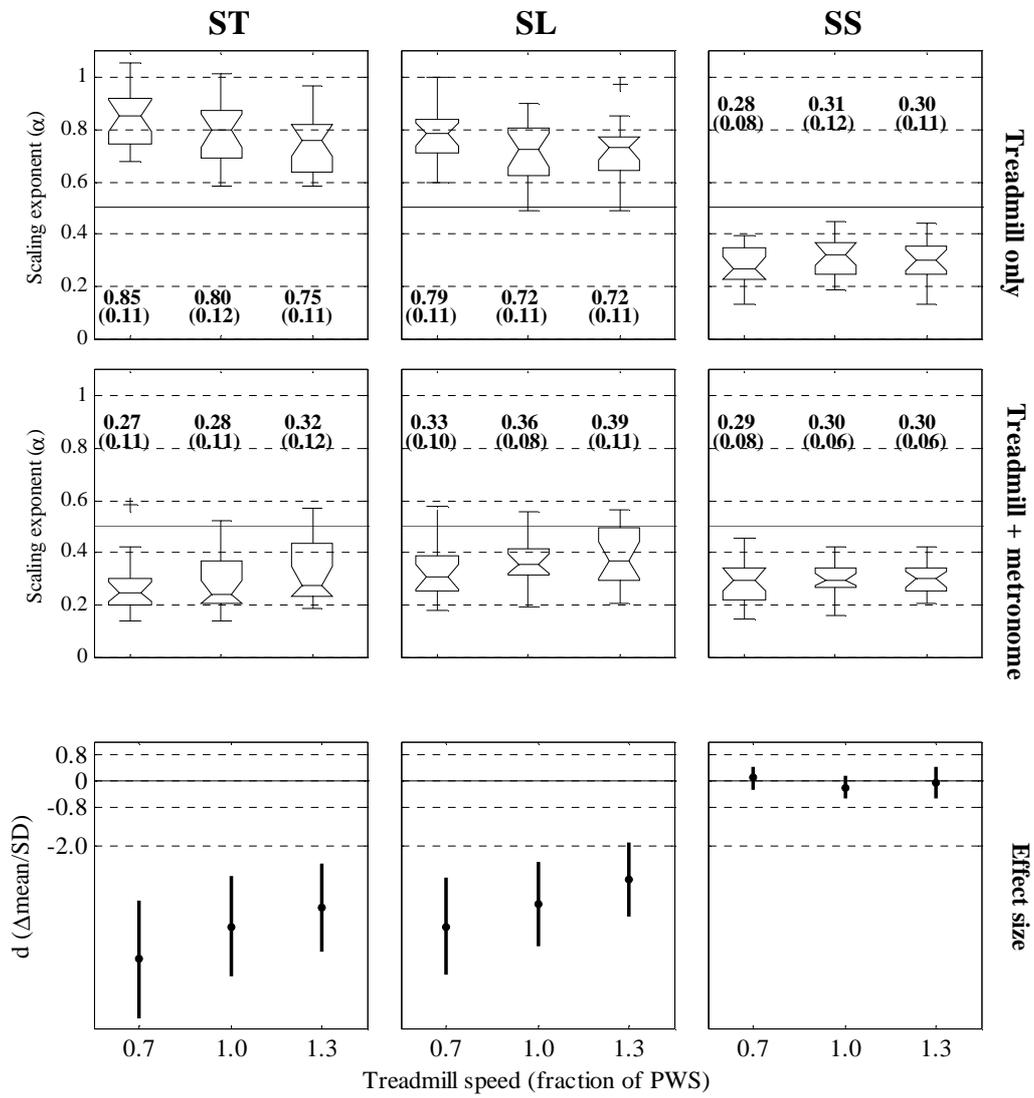

*Figure 4*

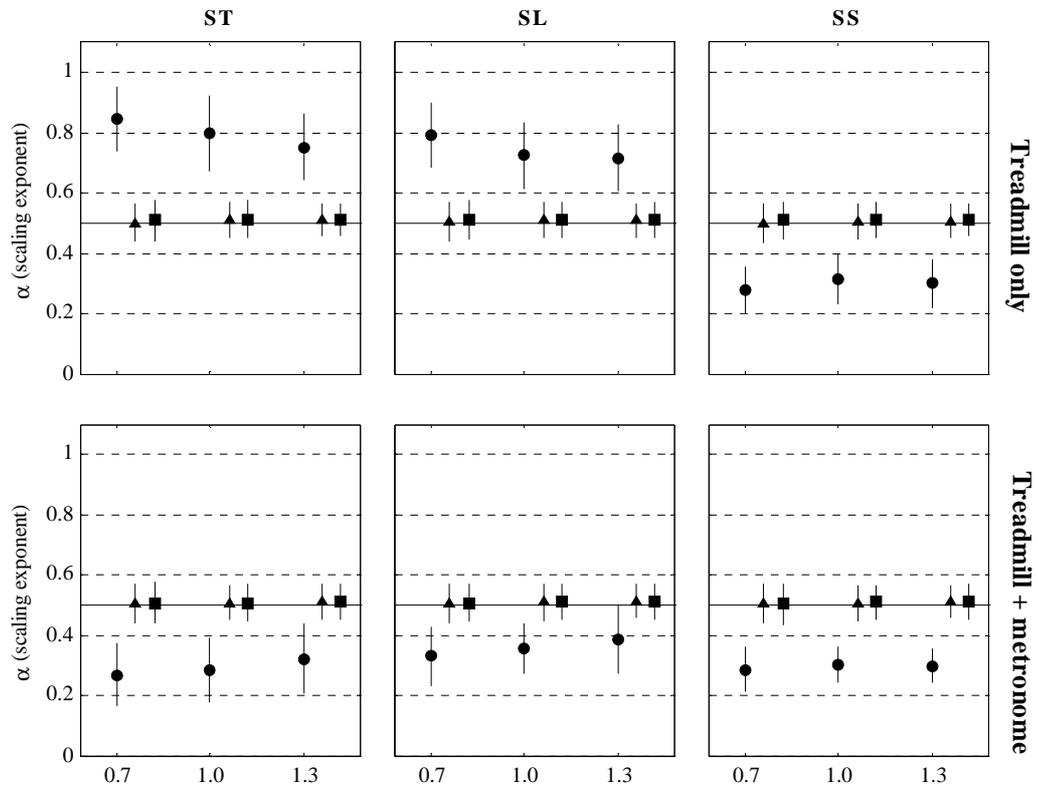

*Figure 5*

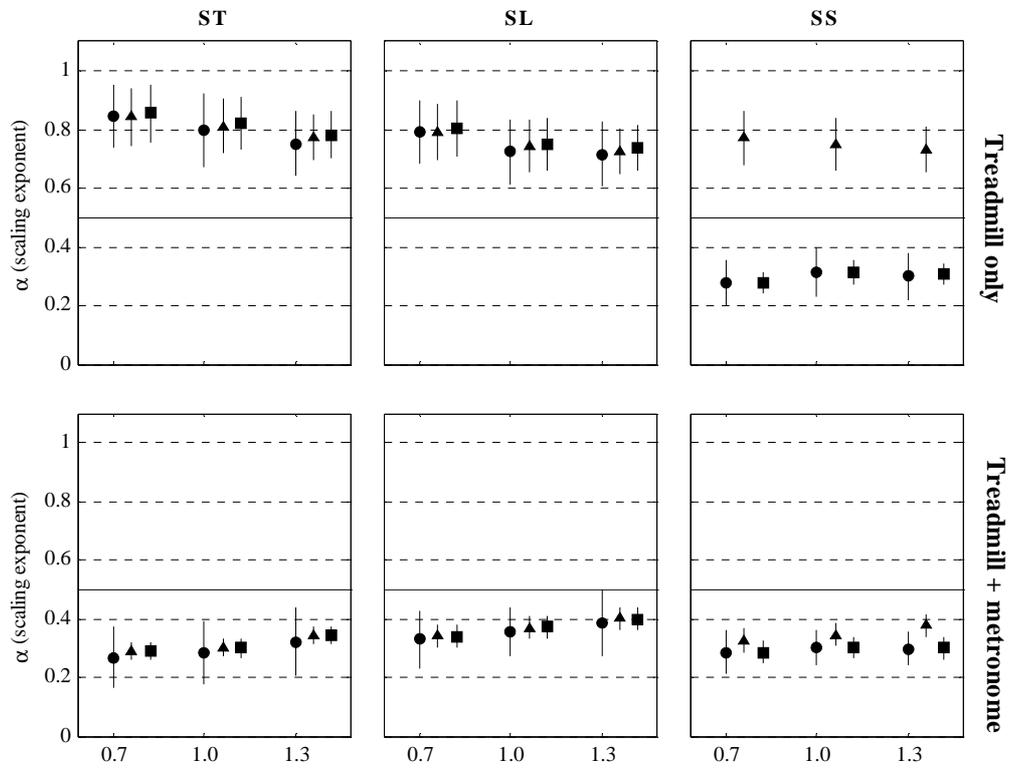